\documentclass[a4paper,11pt]{article}
\usepackage{pos}

\definecolor{hotpink}{rgb}{1,0.27.0.635}
\definecolor{codegray}{rgb}{0.5,0.5,0.5}
\definecolor{codepurple}{rgb}{0.58,0,0.82}
\definecolor{backcolour}{rgb}{0.95,0.95,0.92}
\definecolor{nicegreen}{rgb}{0.,0.5,0.}
\definecolor{darkblue}{rgb}{0.0,0,0.5}

\title{CT25: Progress toward next-generation PDFs for precision phenomenology at the LHC}

\author[a]{A.~Ablat}
\author[b]{A.~Courtoy}
\author[a,c]{S.~Dulat}
\author[c]{Y. Fu}
\author[d]{M.~Guzzi}
\author*[e]{T.~J.~Hobbs}
\author[c]{J.~Huston}
\author[c]{K.~Mohan}
\author[c]{P.~Nadolsky}
\author[c]{M. Ponce-Chavez}
\author[c]{D.~Stump}
\author[c]{K.~Xie}
\author[c]{C.-P.~Yuan}

\emailAdd{tim@anl.gov}
\affiliation[a]{School of Physics Science and Technology, Xinjiang University, Urumqi, Xinjiang 830046 China}
\affiliation[b]{UNAM, Apartado Postal 20-364, 01000 Ciudad de México, Mexico}
\affiliation[c]{Department of Physics \& Astronomy, Michigan State University,
East Lansing, MI 48824, USA}
\affiliation[d]{Kennesaw State University, Kennesaw, GA 30144, USA}
\affiliation[e]{High Energy Physics Division, Argonne National Laboratory, Lemont, IL, USA}

\abstract{
We summarize recent progress toward the next generation of CTEQ-TEA parton
distribution functions, CT25, based on a global NNLO analysis that
incorporates a significant sample of newly included LHC data.
We present a baseline fit within the forthcoming full CT25 fit, which includes
new Drell-Yan, top-pair, and inclusive-jet data at 8 and 13~TeV, and exhibits
non-trivial pulls on the high-$x$ gluon and the flavor structure of the quark sea.
In the context of progress toward CT25, we also summarize several recent and
ongoing studies of the interplay between phenomenological PDFs
and lattice-QCD calculations, simultaneous extractions of
$\alpha_s(M_Z)$ within the CT framework, and an expanded program of
uncertainty quantification that treats parametrization dependence as an
explicit source of epistemic uncertainty, among other issues.
We also briefly highlight CT efforts to understand the effects of partial 
implementations of N$^3$LO corrections into PDF fits, which include benchmark calculations for Higgs and vector-boson processes. We comment on the
implications of recent improvements to the CT analysis for precision phenomenology at the LHC and future facilities.
}

\FullConference{XXXII International Workshop on Deep Inelastic Scattering and Related Subjects (DIS2025)\\
24-28 March, 2025\\
Cape Town, South Africa\\}

\begin{document}
\begin{flushright}
    ANL-201496, MSUHEP-25-026
\end{flushright}

\maketitle

\section{Introduction}
\label{sec:intro}
The CT family of global PDF fits, alongside other QCD analysis efforts, has provided
essential information shaping baseline predictions for LHC precision phenomenology in
recent years.
However, the steady influx of high-precision LHC measurements, the emergence
of lattice-QCD constraints of growing quality to constrain PDFs, increasing accuracy
in perturbative calculations, and novel methods in PDF parameter space exploration and numerical model building have all motivated a new generation of CT analyses.
This newest PDF set, of which we announce a baseline in these brief proceedings, we 
denote collectively as CT25; CT25 is the successor to the previous CT18~\cite{Hou:2019efy} main release and absorbs several modifications and lessons
from dedicated physics studies since the release of CT18.
The DIS~2025 talk in Cape Town, on which these proceedings are based reported on
a series of targeted studies addressing
uncertainty quantification, possible PDF-lattice synergies, and investigations of higher-order QCD and QED corrections in PDF fits;
in addition, we presented the status of a preliminary CT25 fit qualitatively similar to the baseline fit appearing below.
Regarding the latter CT25 analysis and associated studies, we emphasize several priorities, among them:
the inclusion of key LHC data sets (Drell-Yan, $t\bar t$, inclusive jets, and dijets) at NNLO in a consistent global fit;
Investigation of the impacts of lattice-QCD pseudo-PDF results on the gluon and strange sectors;
simultaneous fits of PDFs and $\alpha_s(M_Z)$, with a careful treatment of global and dynamical tolerances;
quantification of parametrization and methodological uncertainties using both flexible analytic forms ({\it e.g.}, B\'ezier curves) and machine-learning-based approaches;
benchmarking partial-N$^3$LO (pN$^3$LO) corrections in key Higgs and electroweak channels against existing NNLO predictions.
While we will not elaborate on the activities above in great detail in these brief proceedings, we will quickly summarize them in Sec.~\ref{sec:dedicated}
as a means of providing context to the forthcoming CT25 release.
Regarding CT25, we highlight the current status in Sec.~\ref{sec:CT25}
below, while also pointing readers to a novel set of PDF grids
containing a main set of parton densities to appear in the full
CT25 release.


\section{Recent and ongoing dedicated studies in the CT family}
\label{sec:dedicated}
The CT18 PDF analysis was the vanguard of a series
of dedicated studies in collider phenomenology exploring the downstream implications of the PDFs --- an exercise initiated by the CT18 study itself, which explored the PDF dependence
of Higgs, $t\bar{t}$, and electroweak cross sections.
Subsequent studies investigated many subjects, including
nonperturbative charm~\cite{Guzzi:2022rca}, simultaneous fits alongside SMEFT~\cite{Gao:2022srd},
QED PDFs~\cite{Xie:2021equ}, and implications for high-energy neutrino scattering~\cite{Xie:2023suk},
among other topics. On the side of PDF methodology, a number
of investigations explored novel approaches to PDF parametrization, such as
Ref.~\cite{Kotz:2023pbu}.

At the DIS25 Meeting, a number of more recent (as well as in-progress) studies
were also discussed, many of them building on aspects of the calculations noted
above.
In addition to the detailed investigation of new data sets for CT25 noted
in Sec.~\ref{sec:CT25}, these included a recent exploration of the PDF-lattice
connection as explored in Ref.~\cite{Ablat:2025xzm}, which analyzed lattice
pseudo-PDF output with a particular focus on the interplay with inclusive-jet
data.

{\bf Parametrization and uncertainty studies}. 
Recent CT effort concentrated on additional studies of the PDF parametrization and related questions on model uncertainties.
In the standard Hessian approach, the \(\Delta\chi^2 = 1\) criterion generally underestimates PDF uncertainties in part because it assumes a linear dependence of observables on PDF parameters defined at the initial scale \(Q_0\). To go beyond this approximation, nonlinear effects can be systematically included via a Taylor expansion~\cite{{Hou:2016sho}}. In Ref.~\cite{Zhan:2024tic}, we introduced an improved Hessian method that incorporates additional error sets from second-order derivatives and a nonlinear uncertainty term quantifying deviations from the linear approximation. Studies with pseudodata show that nonlinear uncertainties can be substantially larger than standard Hessian estimates, especially at low or high \(x\) and for poorly constrained flavors such as sea quarks. Moreover, as demonstrated in the Appendix of Ref.~\cite{Zhan:2024tic}, the inclusion of second-order terms consistently increases PDF-induced uncertainties.
%
%
We note that, in some CT25 fits, we use a dynamical tolerance (D-TOL) prescription which ultimately yields PDF error bands that are comparable, and smaller in some parameter directions, as those obtained with the CT18 two-tier prescription.

Ref.~\cite{Kotz:2025une} extended initial studies based on the B\'ezier curve approach to PDF parametrization in a public release of the Fant\^omas code, including but standalone modules as well as \texttt{xFitter} integrations. This framework unlocks parametrizations of effectively arbitrary flexibility using a tunable decomposition of the PDF starting-scale distribution in terms of carrier and modulator functions, as deployed in further recent studies of the pion structure~\cite{Kotz:2025lio}.
Parallel and complementary developments presented at DIS25 involve the machine-learning models like Dirichlet-Prior Networks~\cite{Kriesten:2024ist} for the separation and classification of BSM signatures from PDFs (including epistemic and aleatoric uncertainties), as well as guided backpropagation methods~\cite{Kriesten:2024are} for emulating PDF analyses while connecting theory assumptions to $x$-dependent features.
Additional efforts toward the classification of PDF solutions are being developed~\cite{Courtoy:2025ppd} based on information-theoretic methods.

\begin{figure}[t]
  \centering
  \raisebox{0.9ex}{%
  \includegraphics[width=0.52\linewidth]{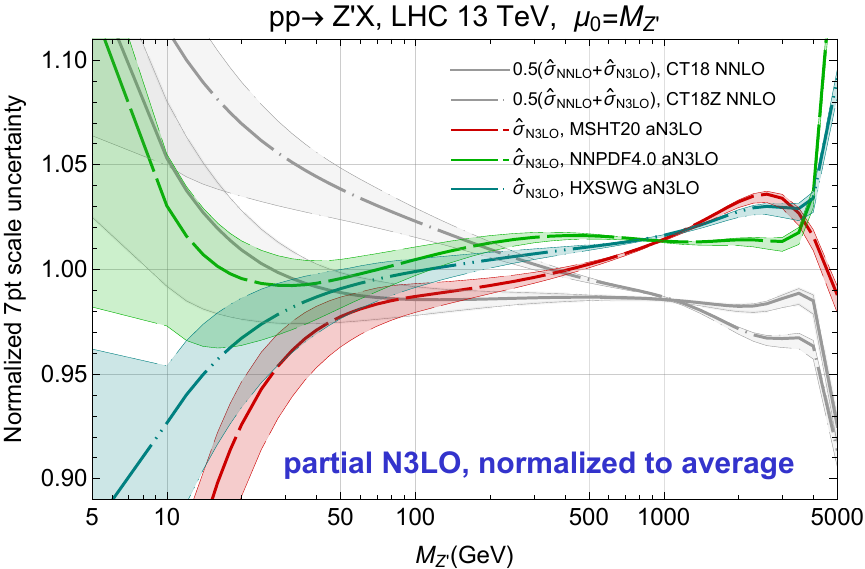}
  }\hfill
  \includegraphics[width=0.45\linewidth]{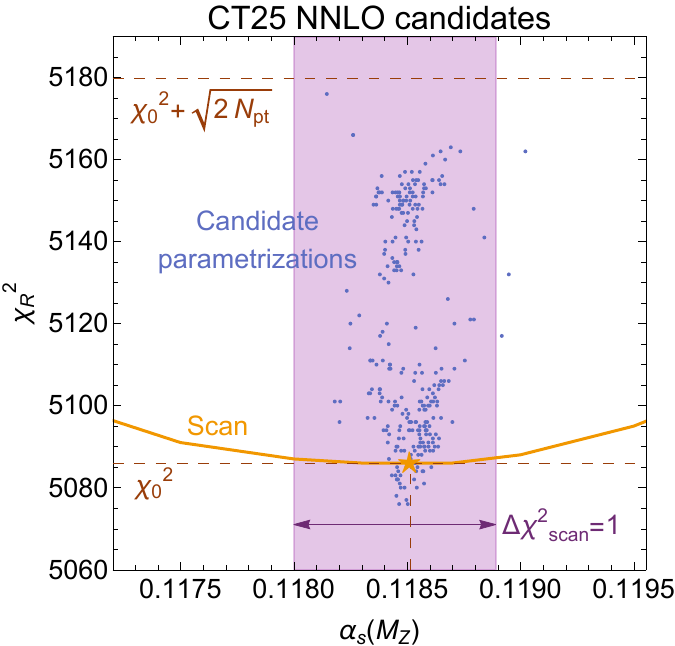}
\caption{({\it Left}) A plot of the $pp\to Z^\prime X$ total cross section {\it vs}.~$M_{Z^\prime}$ in several recent partial N3LO calculations, including an ``NNLO+'' prescription, convoluted with several NNLO and aN3LO PDFs; we note ``HXSWG aN3LO'' refers to the combination set of Ref.~\cite{Cridge:2024icl}. ({\it Right}) $\chi^2(\alpha_s)$ profile from the scan of the preferred CT25 PDF parametrization form and different alternative parametrizations. $\chi^2_0$ represents the $\chi^2$ minimum relative to the best fit.}
  \label{fig:CT-alphas}
\end{figure}

{\bf Perturbative accuracy}. The CT group has also been engaged in multiple efforts to improve the perturbative accuracy of PDF extractions while quantifying the phenomenological implications. In the case of QCD accuracy, Ref.~\cite{Guzzi:2024can} presented a streamlined method to implement the SACOT-MPS general-mass factorization scheme across arbitrary LHC processes using subtraction charm and bottom PDFs published as LHAPDF interpolation tables. These PDFs were applied to calculate the mass-dependent terms in $Z+b$ production at the LHC.
Meanwhile, CT also presented an upcoming exploration of partial implementations of N3LO corrections; this study finds the effect of shifting from a consistent NNLO global analysis ({\it i.e.}, NNLO PDFs fitted with complete NNLO hard matrix elements) to an analysis implementing some N3LO pieces can leave sizable differences among theoretical predictions which are traceable to other systematics of the PDF fit and which are unlikely to be reduced by partial N3LO implementations. A visualization of this can be in the left panel of Fig.~\ref{fig:CT-alphas}, which plots the $pp\to Z^\prime X$ total cross section as a function of the $Z^\prime$ mass under several pN3LO prescriptions.
Along these lines, efforts to improve QCD accuracy motivate parallel developments of QED and electroweak corrections.
Ref.~\cite{Cooper-Sarkar:2025sqw} examined Higgs-sector phenomenology in the presence of QED PDFs based on the LUX
formalism, finding good stability in both total and differential Higgs cross sections given modern QED PDFs, including in the CT family.

\pagebreak

{\bf Strong coupling determination}.
The CT25 global analysis determines $\alpha_s(M_Z)$ at NNLO in QCD using an improved methodology and an expanded data set as compared to the CT18 study~\cite{Hou:2019efy}, including high-luminosity LHC Run-2 measurements. A key advance is the use of multiple methods to derive the central value and uncertainty of $\alpha_s(M_Z)$, exploring sensitivity to PDF parametrizations, data selection, and treatment of systematic errors.
The resulting $\alpha_s(M_Z)$ range is less sensitive to the definition of the PDF uncertainty (based on tolerance criteria~\cite{Kovarik:2019xvh}), as compared to previous combined PDF+$\alpha_s$ fits.
The consistency of this range of $\alpha_s(M_Z)$ values with the global dataset at NNLO is established by using different methods, such as the generation of a $\chi^2(\alpha_s)$ profile via a single global fit, the comparison of $\alpha_s(M_Z)$ values from multiple fits employing independent PDF parametrizations, and the combination of experimental constraints via varied criteria to evaluate the uncertainty. The final value 
$\alpha_s(M_Z) = 0.1183 \pm 0.00225$, is derived by synthesizing these extraction methods using different statistical models for the combination of the results. We find that the impact from many [${\cal O}(300)$] alternative PDF parametrizations on the $\chi^2$-profile is small and compatible with the $\Delta\chi^2 =1$ criterion (see Fig.~\ref{fig:CT-alphas}), while the dependence of the relative uncertainty on other tolerance criteria as well as on different treatments of the systematic uncertainties is substantial, due to disagreements among data sets in the baseline. 

Having highlighted select recent and forthcoming developments within the CT family,
we briefly summarize the status of the CT25 PDF release below, drawing
attention to the newly added data sets, theoretical considerations, and PDF behaviors
expected of the full CT25 analysis.


\section{The upcoming CT25 main release}
\label{sec:CT25}
\begin{figure}[t]
  \centering
  \includegraphics[width=0.5\linewidth]{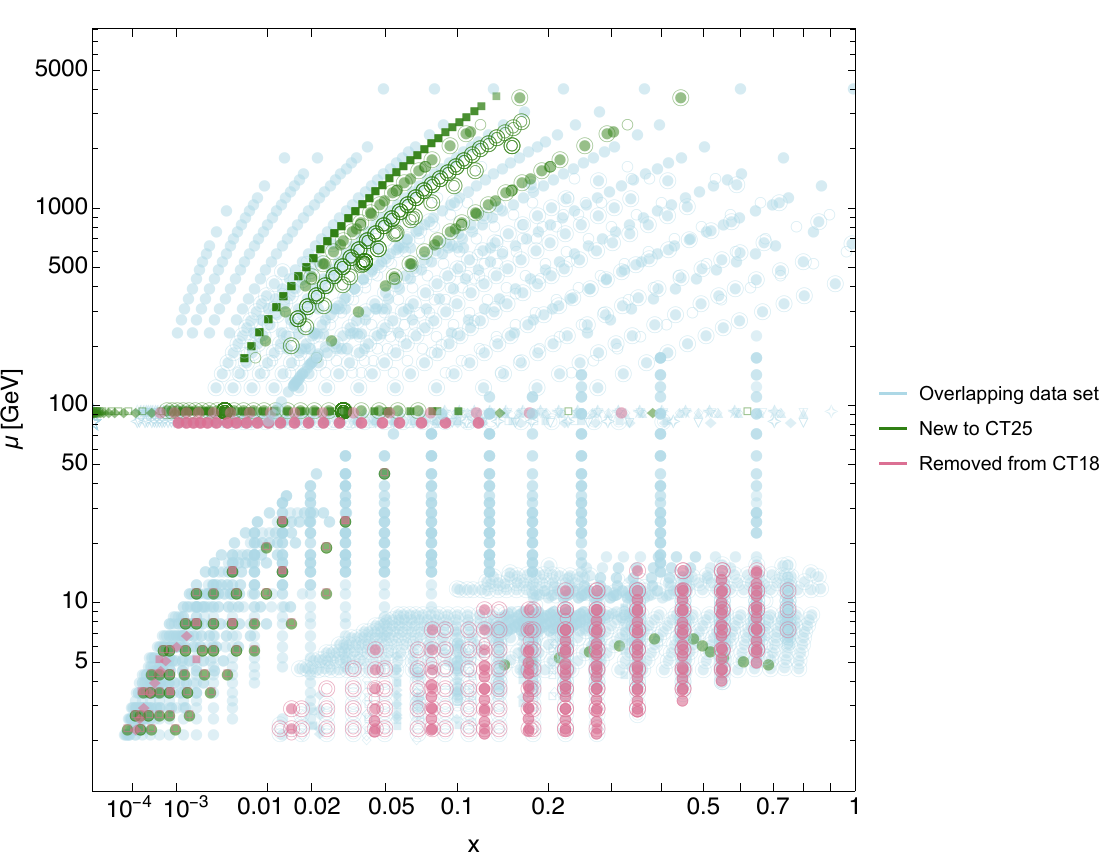}
    \includegraphics[width=0.47\linewidth]{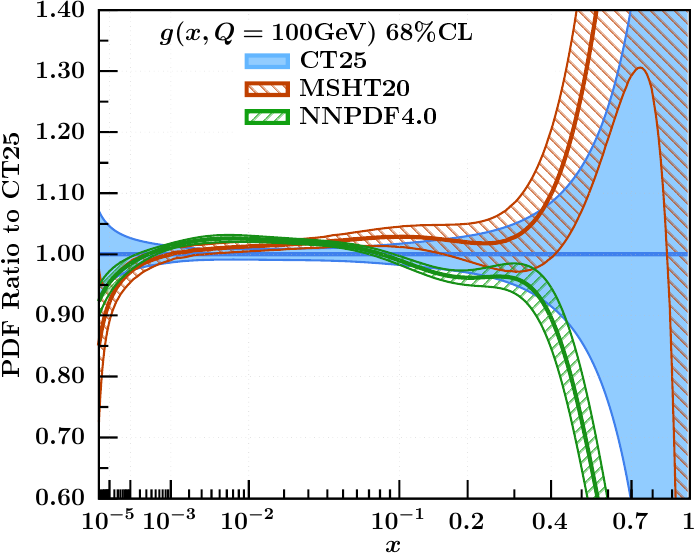}
\caption{Left: The distribution of $x$ and ~$\mu$ values of data points in the upcoming
	CT25 global PDF analysis. We show the newly added data points in CT25 in green, 
    in contrast to the removed ones, shown in red.
	Points common to both the CT25 and CT18 baselines are represented in blue. Right: Gluon uncertainties from CT25, MSHT20, and NNPDF4.0 error PDF ensembles.}
  \label{fig:CT-data}
\end{figure}

The upcoming CT25 release incorporates several new experimental data from the LHC, including several Drell-Yan pair-production experiments at 5, 8, and 13 TeV from ATLAS, CMS, and LHCb; several dozen points on 13 TeV $t\bar{t}$ production cross section, differential in both top-pair rapidity, $y_{t\bar{t}}$ and invariant mass, $m_{t\bar{t}}$, from ATLAS and CMS; and 531 data points from inclusive-jet cross-section
measurements at 13 TeV from ATLAS and CMS. The latter inclusive jet measurements were investigated in conjunction with dijet measurements in a separate CT study~\cite{Ablat:2024uvg}, which found the perturbative scale variations of the inclusive jet data to be under comparatively better control.
The kinematical coverage of these newly introduced experiments is shown in Fig.~\ref{fig:CT-data}, indicated by green points. The distribution of points in
Fig.~\ref{fig:CT-data} highlights the increased constraining power of the upcoming
CT25 analysis, especially with respect to the low-$x$ region over a range of
factorization scales, $\mu$.
We also note that the PDF sensitivity of the data included in the CT25 baseline
is further supported by default CT tools, including the $L_2$-sensitivity method,
which suggests significant pulls for various data sets. These include, {\it e.g.}, the CMS 13 TeV top-pair data
(differential in both $m_{t\bar{t}}$ and $y_{t\bar{t}}$), which show sensitivity to the gluon PDF in CT25 candidate fits, as well as the inclusive jet data. Meanwhile, fixed-target Drell-Yan information from SeaQuest add additional information on ${\bar d}/{\bar u}$ at $x> 0.3$.

While a full dissection of the PDF impacts of these new data and corresponding
adjustments to the theory is beyond the scope of these short proceedings, 
the right panel of Fig.~\ref{fig:CT-data} 
illustrates the comparison of uncertainties on gluon PDFs from the nominal CT25, MSHT20, and NNPDF4.0 NNLO error ensembles. CT25 is compatible with CT18A -- its progenitor baseline PDF -- while having a mildly higher $u$ and $d$, as well as mildly lower $g$ PDFs at $x=0.01-0.1$ relevant for Higgs and other LHC measurements. The shown CT25 PDF ensemble is determined with one selection of PDF functional forms, using $\alpha_s(M_Z)=0.118$ and a CT-developed dynamical tolerance prescription for the PDF uncertainties. As in the case of CT18, other PDF ensembles in the CT25 family will be released to account for variations of $\alpha_s$, choices of the fitted data sets, and the maximal number of flavors, and at LO and NLO. Using a variety of functional forms, we dissect the sensitivity of PDFs to the flavor decomposition of $\bar u$, $\bar d$, $\bar s$ at $x>0.2$ and assumptions about the low-$x$ behavior of the gluon.

Finally, we point out that the nominal CT25 set presented in brief above, based on the CT realization of the dynamical tolerance prescription, is now available at the CT public page~\cite{CTwebsite}. Comprehensive details on the CT25 PDFs and their implementation will be discussed in the forthcoming CT25 long paper.


\section{Conclusions and outlook}
\label{sec:conc}
The CT25 fit, as illustrated by our initial baseline release, demonstrates that a carefully curated ensemble of LHC measurements, including from Run-2, can enhance PDF
precision in phenomenologically important regions while preserving
global consistency with earlier data.


\section*{Acknowledgments}
 
AC was supported by the UNAM Grant No.~DGAPA-PAPIIT IN102225. 
 MG was supported by the U.S.~National Science Foundation under Grant No.~PHY-2412071. 
The work of T.~J.~Hobbs at Argonne National Laboratory was supported by the U.S.~Department of Energy, Office of Science, under Contract No. DE-AC02-06CH11357. 
The work of CPY and KX was supported by the U.S. National Science Foundation under
Grant No.~PHY-2310291. 
KM and KX were also supported by the U.S. National Science Foundation under
Grant No.~PHY-PHY-2310497. 
PMN and SD are grateful for support from the Wu-Ki Tung Endowed Chair in particle physics. 
This work utilized resources from high-performance computing clusters at SMU M3, MSU HPCC, KSU HPC, and ANL LCRC.

\bibliographystyle{JHEP}
\bibliography{references}

\providecommand{\href}[2]{#2}\begingroup\raggedright\begin{thebibliography}{10}

\bibitem{Hou:2019efy}
T.-J. Hou et~al., \emph{{New CTEQ global analysis of quantum chromodynamics
  with high-precision data from the LHC}},
  \href{https://doi.org/10.1103/PhysRevD.103.014013}{\emph{Phys. Rev. D}
  {\bfseries 103} (2021) 014013}
  [\href{https://arxiv.org/abs/1912.10053}{{\ttfamily 1912.10053}}].

\bibitem{Guzzi:2022rca}
M.~Guzzi, T.~J. Hobbs, K.~Xie, J.~Huston, P.~Nadolsky and C.~P. Yuan,
  \emph{{The persistent nonperturbative charm enigma}},
  \href{https://doi.org/10.1016/j.physletb.2023.137975}{\emph{Phys. Lett. B}
  {\bfseries 843} (2023) 137975}
  [\href{https://arxiv.org/abs/2211.01387}{{\ttfamily 2211.01387}}].

\bibitem{Gao:2022srd}
J.~Gao, M.~Gao, T.~J. Hobbs, D.~Liu and X.~Shen, \emph{{Simultaneous CTEQ-TEA
  extraction of PDFs and SMEFT parameters from jet and $ t\overline{t} $
  data}}, \href{https://doi.org/10.1007/JHEP05(2023)003}{\emph{JHEP} {\bfseries
  05} (2023) 003} [\href{https://arxiv.org/abs/2211.01094}{{\ttfamily
  2211.01094}}].

\bibitem{Xie:2021equ}
{\scshape CTEQ-TEA} collaboration, \emph{{Photon PDF within the CT18 global
  analysis}}, \href{https://doi.org/10.1103/PhysRevD.105.054006}{\emph{Phys.
  Rev. D} {\bfseries 105} (2022) 054006}
  [\href{https://arxiv.org/abs/2106.10299}{{\ttfamily 2106.10299}}].

\bibitem{Xie:2023suk}
{\scshape CTEQ-TEA} collaboration, \emph{{High-energy neutrino deep inelastic
  scattering cross sections}},
  \href{https://doi.org/10.1103/PhysRevD.109.113001}{\emph{Phys. Rev. D}
  {\bfseries 109} (2024) 113001}
  [\href{https://arxiv.org/abs/2303.13607}{{\ttfamily 2303.13607}}].

\bibitem{Kotz:2023pbu}
L.~Kotz, A.~Courtoy, P.~Nadolsky, F.~Olness and M.~Ponce-Chavez,
  \emph{{Analysis of parton distributions in a pion with B{\'e}zier
  parametrizations}},
  \href{https://doi.org/10.1103/PhysRevD.109.074027}{\emph{Phys. Rev. D}
  {\bfseries 109} (2024) 074027}
  [\href{https://arxiv.org/abs/2311.08447}{{\ttfamily 2311.08447}}].

\bibitem{Ablat:2025xzm}
A.~Ablat, S.~Dulat, T.-J. Hou, H.-W. Lin, K.~Xie and C.~P. Yuan, \emph{{Impact
  of lattice gluon dataset on CTEQ-TEA global PDFs}},
  \href{https://arxiv.org/abs/2502.10630}{{\ttfamily 2502.10630}}.

\bibitem{Hou:2016sho}
T.-J. Hou et~al., \emph{{Reconstruction of Monte Carlo replicas from Hessian
  parton distributions}},
  \href{https://doi.org/10.1007/JHEP03(2017)099}{\emph{JHEP} {\bfseries 03}
  (2017) 099} [\href{https://arxiv.org/abs/1607.06066}{{\ttfamily
  1607.06066}}].

\bibitem{Zhan:2024tic}
W.~Zhan, S.~Yang, M.~Liu, L.~Han, D.~Stump and C.~P. Yuan, \emph{{Improved
  Hessian method in global analysis of parton distribution functions}},
  \href{https://doi.org/10.1103/858l-2x47}{\emph{Phys. Rev. D} {\bfseries 112}
  (2025) 074028} [\href{https://arxiv.org/abs/2411.11645}{{\ttfamily
  2411.11645}}].

\bibitem{Kotz:2025une}
L.~Kotz, A.~Courtoy, T.~Hobbs, P.~Nadolsky, F.~Olness, M.~Ponce-Chavez et~al.,
  \emph{Fantômas unconfined: global qcd fits with bézier parameterizations},
  \href{https://doi.org/https://doi.org/10.1016/j.cpc.2025.109969}{\emph{Computer
  Physics Communications} (2025) 109969}.

\bibitem{Kotz:2025lio}
L.~Kotz, A.~Courtoy, P.~Nadolsky and M.~Ponce-Chavez, \emph{{Epistemic and
  nuclear uncertainties for the parton distributions of the pion}},
  \href{https://doi.org/10.1103/h2vn-1wxp}{\emph{Phys. Rev. D} {\bfseries 112}
  (2025) L071502} [\href{https://arxiv.org/abs/2505.13594}{{\ttfamily
  2505.13594}}].

\bibitem{Kriesten:2024ist}
B.~Kriesten and T.~J. Hobbs, \emph{{Anomalous electroweak physics unraveled via
  evidential deep learning}},
  \href{https://doi.org/10.1140/epjc/s10052-025-14501-6}{\emph{Eur. Phys. J. C}
  {\bfseries 85} (2025) 883}
  [\href{https://arxiv.org/abs/2412.16286}{{\ttfamily 2412.16286}}].

\bibitem{Kriesten:2024are}
B.~Kriesten, J.~Gomprecht and T.~J. Hobbs, \emph{{Explainable AI classification
  for parton density theory}},
  \href{https://doi.org/10.1007/JHEP11(2024)007}{\emph{JHEP} {\bfseries 11}
  (2024) 007} [\href{https://arxiv.org/abs/2407.03411}{{\ttfamily
  2407.03411}}].

\bibitem{Courtoy:2025ppd}
A.~Courtoy and A.~Ibsen, \emph{{Information Criteria for Selecting Parton
  Distribution Function Solutions}},
  \href{https://arxiv.org/abs/2511.07518}{{\ttfamily 2511.07518}}.

\bibitem{Guzzi:2024can}
M.~Guzzi, P.~Nadolsky, L.~Reina, D.~Wackeroth and K.~Xie, \emph{{General mass
  variable flavor number scheme for Z boson production in association with a
  heavy quark at hadron colliders}},
  \href{https://doi.org/10.1103/PhysRevD.110.114030}{\emph{Phys. Rev. D}
  {\bfseries 110} (2024) 114030}
  [\href{https://arxiv.org/abs/2410.03876}{{\ttfamily 2410.03876}}].

\bibitem{Cooper-Sarkar:2025sqw}
A.~M. Cooper-Sarkar, T.~Cridge, T.~J. Hobbs, J.~Huston, P.~Nadolsky,
  M.~Ponce-Chavez et~al., \emph{{QED-enhanced PDF implications for the Higgs
  sector}},  \href{https://arxiv.org/abs/2508.06603}{{\ttfamily 2508.06603}}.

\bibitem{Kovarik:2019xvh}
K.~Kova{\v{r}}{\'\i}k, P.~M. Nadolsky and D.~E. Soper, \emph{{Hadronic
  structure in high-energy collisions}},
  \href{https://doi.org/10.1103/RevModPhys.92.045003}{\emph{Rev. Mod. Phys.}
  {\bfseries 92} (2020) 045003}
  [\href{https://arxiv.org/abs/1905.06957}{{\ttfamily 1905.06957}}].

\bibitem{Cridge:2024icl}
T.~Cridge et~al., \emph{{Combination of aN$^3$LO PDFs and implications for
  Higgs production cross-sections at the LHC}},
  \href{https://doi.org/10.1088/1361-6471/adde78}{\emph{J. Phys. G} {\bfseries
  52} (2025) 6} [\href{https://arxiv.org/abs/2411.05373}{{\ttfamily
  2411.05373}}].

\bibitem{Ablat:2024uvg}
{\scshape CTEQ-TEA} collaboration, \emph{{Impact of LHC precision measurements
  of inclusive jet and dijet production on the CTEQ-TEA global PDF fit}},
  \href{https://doi.org/10.1103/PhysRevD.111.036033}{\emph{Phys. Rev. D}
  {\bfseries 111} (2025) 036033}
  [\href{https://arxiv.org/abs/2412.00350}{{\ttfamily 2412.00350}}].

\bibitem{CTwebsite}
\protect{CT25 PDFs}. \url{https://cteq-tea.gitlab.io/project/00pdfs/}.

\end{thebibliography}\endgroup
\end{document}